\documentclass[a4paper]{article}

\usepackage{INTERSPEECH2015}

\usepackage[utf8]{inputenc}
\usepackage{graphicx}
\usepackage{amssymb,amsmath,bm}
\usepackage{textcomp}
\usepackage{hyperref}
\usepackage{todonotes}

\ninept

\title{Using Deep Learning for Detecting Spoofing Attacks on Speech Signals}

\makeatletter
\def\name#1{\gdef\@name{#1\\}}
\makeatother
\name{{\em Alan Godoy$^{1,2}$, Flávio Simões$^1$, José Augusto Stuchi$^1$, Marcus de Assis Angeloni$^1$,} \\
{\em Mário Uliani$^1$, Ricardo Violato$^1$}}

\address{$^1$CPqD Foundation, Campinas, Brazil \\
  $^2$University of Campinas, Campinas, Brazil \\
  {\small \tt \{amello,simoes,jastuchi,massis,uliani,rviolato\}@cpqd.com.br}
}

\begin{document}

  \maketitle
  \begin{abstract}
    It is well known that speaker verification systems are subject to spoofing attacks. The Automatic Speaker Verification Spoofing and Countermeasures Challenge -- ASVSpoof2015 -- provides a standard spoofing database, containing attacks based on synthetic speech, along with a protocol for experiments. This paper describes CPqD's systems submitted to the ASVSpoof2015 Challenge, based on deep neural networks, working both as a classifier and as a feature extraction module for a GMM and a SVM classifier. Results show the validity of this approach, achieving less than 0.5\% EER for known attacks.
  \end{abstract}

  \section{Introduction}
  \label{sec:introduction}

    Biometric spoofing is usually described as a direct attack perpetrated against a biometric authentication system by presenting it a fake (forged or copied) biometric sample. Anti-spoofing refers, therefore, to countermeasures designed to detect and prevent these attacks \cite{evans2013spoofing}.

    In the last few years, many studies have shown that even state-of-the-art automatic speaker verification (ASV) systems are vulnerable to such attacks, which can be based on a variety of techniques, including voice conversion, speech synthesis, artificial signals, impersonation, and replay \cite{evans2013spoofing}. Although most of these studies proposes countermeasures too, they usually are based on prior knowledge about the attack method, what is clearly unrepresentative of real world scenarios. Additionally, each one is also based on its own database, protocol and metrics, making it difficult to perform a proper analysis of results and restricting fair comparison among them \cite{wu2015asvspoof}.

    The recent Automatic Speaker Verification Spoofing and Countermeasures Challenge, ASVSpoof2015\footnote{\url{http://www.spoofingchallenge.org/}}, which focused on spoofing attacks based on synthetic speech, provided the first standard spoofing database along with a protocol for experiments. Differently from previous works, 10 different voice conversion and speech synthesis algorithms were used to generate the database, but only 5 of them were known in advance in order to train spoofing detection algorithms \cite{wu2014asvspoof}. This paper describes the systems based on neural networks submitted to the challenge and analyze the obtained results.

    Deep Neural Networks (DNN) have been widely used in a variety of research fields, such as image classification \cite{he2015delving, ioffe2015batch}, natural language processing \cite{collobert2008deep} and information retrieval \cite{huang2013a}. In the speech processing community, DNN have been applied to speech recognition \cite{dahl2012context}, speech synthesis \cite{zen2013deep, kang2013mult} and also to speaker recognition \cite{variani2014deep, lei2014novel}.

    One straightforward application of a DNN for spoofing detection is to use it as a classifier, whose input data can be either raw audio \cite{lee2009unsupervised} or features previously extracted from the audio files. A natural choice for audio pre-processing is to use features proven to yield good results in speaker recognition and spoofing detection tasks, such as traditional Mel Frequency Cepstral Coefficients (MFCC) \cite{davis1980comparison} and Modified Group Delay Cepstral Coefficients (MGDCC) \cite{wu2013synthetic}, which have been broadly used not only in combination with neural networks, but also with a handful of other classification algorithms.

    In problems like spoofing detection, a DNN can also be employed as a feature extraction module itself, by means of a bottleneck approach \cite{yu2011improved}. In this case, a network, initially trained for regression or classification, has its final layers removed, and the output of its last remaining layer is used as a new representation of the input data for future classification \cite{lee2009unsupervised}. The network can receive as input a pre-processed feature vector, a high-level full representation of the signal (using, for instance, the Fast Fourier transform) or even the raw audio. In this work, we used the high-level representation approach, as described in Section \ref{sec:method}.

    The paper is organized as follows: Section \ref{sec:neural-networks} presents a brief description of neural networks. Section \ref{sec:method} explains the methods applied. Section \ref{sec:results} presents and discusses results obtained on the ASVspoof2015 challenge. Finally, Section \ref{sec:conclusions} draw some conclusions, as well as points to topics for future research.

  \section{Neural Networks}
  \label{sec:neural-networks}

    The submitted systems are based on a Deep Learning approach. A deep neural network (DNN) is an artificial neural network with more than one hidden neuronal layer between its inputs and outputs \cite{hinton2012deep}. The DNN concept can be implemented using many different architectures, such as Convolutional Neural Networks (CNN) \cite{abdel2013exploring}, Autoencoders \cite{pascal2010stacked}, and Multilayer Perceptrons (MLP) \cite{haykin2009neural}.

    In a Multilayer Perceptron, tipically, each neuron $j$ in a hidden layer $l$ employs a sigmoid function, such as the logistic function or hyperbolic tangent, to map the total input $x^{l}_{j}$, received from the layer $l - 1$, to an output $y^{l}_{j}$, that is sent to the following layer, $l + 1$.

\begin{equation}
        x^{l}_{j} = b^{l}_{j} + \sum_{1 \leq i \leq N^{l - 1}}{w^{l}_{i, j} y^{l - 1}_{i}}
\end{equation}

\begin{equation}
        y^{l}_{j} = logistic(x^{l}_{j})
\end{equation}

\noindent where $N^{l - 1}$ is the number of neurons in layer $l - 1$, $y^{l - 1}_{i}$ is the output of neuron $i$ on previous layer, $w^{l}_{i,j}$ is the connection weight between neuron $i$ from layer $l - 1$ and neuron $j$ from layer $l$, and $b^{l}_{j}$ is the bias of neuron $j$ of the current layer \cite{hinton2012deep}.

    One of the major DNN applications is for multiclass classification problems. In this context, a softmax nonlinear function can be used in the network output layer to convert inputs $x^{out}_{j}$, into a class probability, $p_{j}$:

\begin {equation}
    p_{j} = \frac{\exp (x^{out}_{j})}{\sum_{1 \leq k \leq N^{out}}\exp(x^{out}_{k})}
\end {equation}

\noindent where $N^{out}$ is the number of neurons in the output layer, which is equal to the number of possible classes. In this case, the network output $p_{j}$ will indicate the likelihood of the input fed to the network belonging to the $j$-th class \cite{hinton2012deep}.

  \section{Method}
  \label{sec:method}

  \subsection{Feature Extraction}
  \label{subsec:feature-extraction}

\begin{figure}[!htb]
    \centering
    \includegraphics[scale=0.8]{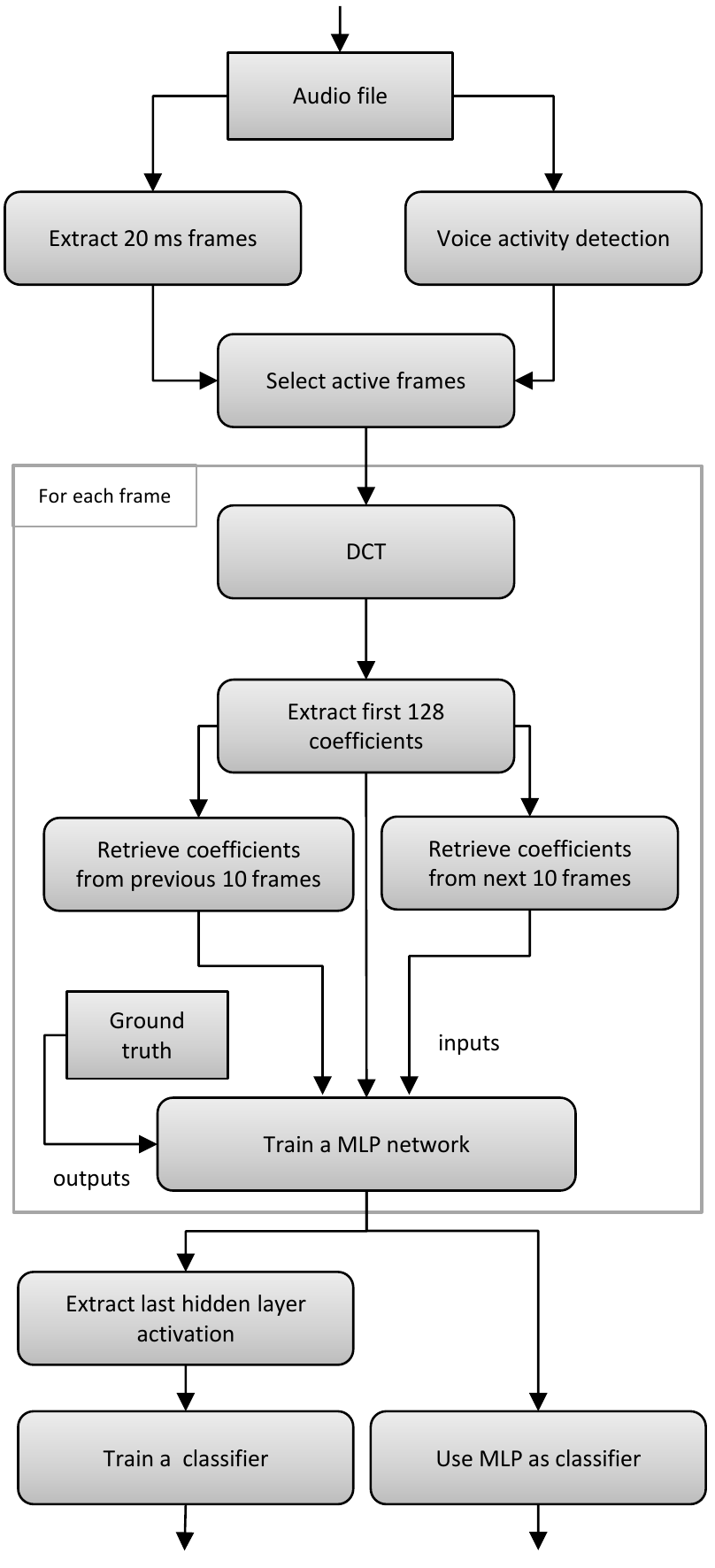}
    \caption{Basic flowchart used for spoofing detection}
    \label{fig:flowchart}
  \end{figure}

    Aiming at detecting if an audio is authentic or not, a deep neural network based on a multilayer perceptron architecture was used as a feature extraction module. In a bottleneck approach, the network output layer is removed and the activations of the last hidden layer neurons are treated as new features for future classification. Figure~\ref{fig:flowchart} shows how audio was processed, from feature extraction to network supervised training.

    Instead of feeding raw signal directly as input to the network, a pre-processing step was performed in order to transform input signals into sequences of feature vectors. This decision was based on preliminary tests, which indicated such a step was able to improve the learning rate and allowed the use of more compact networks. Therefore, each signal file is divided into a sequence of 20 ms consecutive non-overlapping frames. No window function is applied. In parallel, a voice activity detection method based on ITU G.729B \cite{benyassine1997itu} is applied, so each frame is classified as speech/non-speech and only speech frames are preserved.

    Different representations were tested as input for the MLP, including the raw speech frame itself, MFCC, MGDCC and Discrete Fourier Transform (DFT) coefficients. Nevertheless, better results were achieved with the Discrete Cosine Transform (DCT) coefficients. The DCT has the energy compaction property, which concentrates most of the signal information in a few low-frequency components \cite{ahmed1974discrete}. For this reason, the first 128 DCT coefficients are used as feature for each active speech frame.

    In order to avoid loss of long term information that can possibly be used to distinguish spoofing attacks, when an input is presented to the MLP, each central speech frame is surrounded by its ten previous frames and the ten following ones, including silence frames \cite{variani2014deep}. Thus, a vector with 2688 features is used as network input.

    The backpropagation algorithm, in conjunction with the Stochastic Gradient Descent optimization technique \cite{haykin2009neural}, was applied to train the network to classify whether the input represents an authentic (human) or spoofed audio frame. Ground truth consists of a label indicating if the input audio is authentic or belongs to one of five spoofing categories, named S1, S2, S3, S4 or S5 \cite{wu2015asvspoof}.

    Preliminary experiments indicated that using only two classes -- spoofing and human -- as output led to poor perfomance in class S1. One hypothesis is that this could happen because S1 distinguishes from other attacks since it is based on a unit selection algorithm, which concatenates pieces of authentic signal to create a new audio. To deal with this, it was decided to drive the network training towards distinguishing S1 from the other spoofing attacks, increasing the relevance (on network performance) of detecting borders between pieces of authentic speech. Thus, three classes were created, as depicted in Table~\ref{tab:meaning}: authentic human speech (100), S1 spoofing attack (010) and other spoofing attacks (001).

\begin{table}[h]
\caption{\label{tab:mlp_outputs} {\it MLP classes output meanings.}}
\centerline{
\begin{tabular}{|c|c|c|c|}
\hline
\textbf{y0} & \textbf{y1} & \textbf{y2} & \textbf{Meaning}                                                 \\ \hline
1           & 0           & 0           & human                                                            \\ \hline
0           & 1           & 0           & S1 attack                                                        \\ \hline
0           & 0           & 1           & \begin{tabular}[c]{@{}c@{}}S2, S3, S4, S5\\ attacks\end{tabular} \\ \hline
\end{tabular}
}
\label{tab:meaning}
\end{table}

\begin{figure}[!htb]
    \centering
    \includegraphics[scale=0.7]{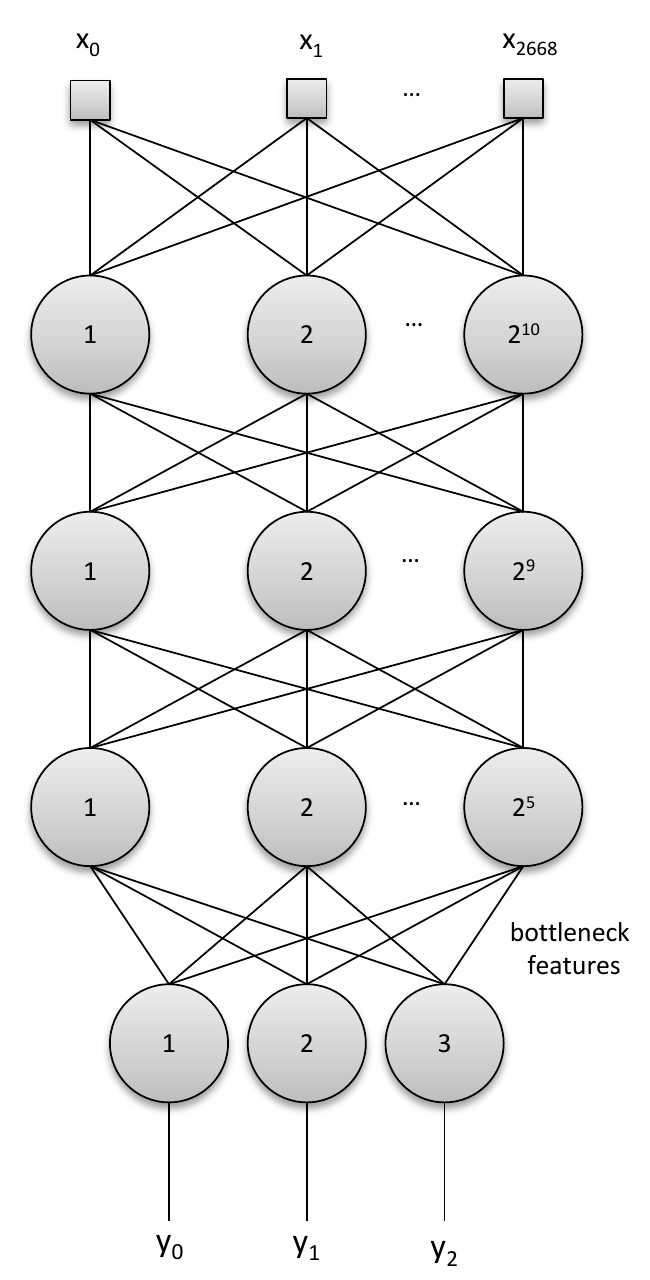}
    \caption{MLP used for feature extraction and classification}
    \label{fig:mlp}
  \end{figure}

    Figure~\ref{fig:mlp} shows the MLP deep architecture used in this paper. 1024 neurons were used in the first  hidden layer, 512 in the second hidden layer and 32 in the last one. The last hidden layer is artificially small in order to create a bottleneck, which compress signal information useful for spoofing classification in a low-dimensional representation \cite{yu2011improved}. Each hidden layer uses the logistic function as activation. The output consists of 3 neurons, each one with softmax activation function, returning a real number between 0 and 1. After finishing the network training, the output layer was removed and the activations of the last hidden layer neurons were used as new output, extracting the bottleneck features, as indicated in Figure~\ref{fig:mlp}.

  \subsection{Classification}
  \label{subsec:classification}

    Three different classifiers were tested: Support Vector Machines (SVM), Gaussian Mixture Models (GMM) and  Multilayer Perceptron. In the cases of the SVM and the GMM classifiers, feature extraction took an additional step. Since each audio file has a different duration and, thus, a different number of frames, feature vectors over all frames were averaged so that each file was represented by a single fixed-size 32-dimensional feature vector \cite{correia2014anti}.

    A SVM classifier \cite{vapnik1995} based on the Radial Basis Function (RBF) kernel was generated. Samples from the training set were computed and used to train the SVM-RBF. All spoofing attacks were considered as a single negative class for training.

    The SVM-RBF classifier parameters $C$ (controls the cost of misclassification on the training data) and $\gamma$ (parameter of a Gaussian kernel to handle nonlinear classification) were tuned by performing grid search with K-fold cross-validation over the train set, using 5 folds. Values of 0.001, 0.01, 0.1, 1.0, 10.0, 100.0, 1000.0 and 10000.0 were searched both for $C$ and $\gamma$. Optimum parameters were chosen aiming at minimizing the average equal error rate (EER) over all 5 folds. After this search, optimum values of $C = 0.1$ and $\gamma = 10$ were found and the SVM-RBF classifier was retrained with the whole training set. SVM-RBF outputs vary in the interval $[0.0, 1.0]$ and represent the likelihood of the test sample belonging to positive class, i.e., authentic speech audio.

    For the GMM based classifier, two GMMs were trained, one with authentic audios and another with spoofed audios. The following number of Gaussian mixtures were tested: 4, 8, 32, 64, 128, 256 and 512, wherein 8 mixtures gave the lowest EER on the development set. The classifier output is given by the log-likelihood ratio of authentic GMM with respect to spoofing GMM.

\begin{figure}[!htb]
    \centering
    \includegraphics[scale=0.35]{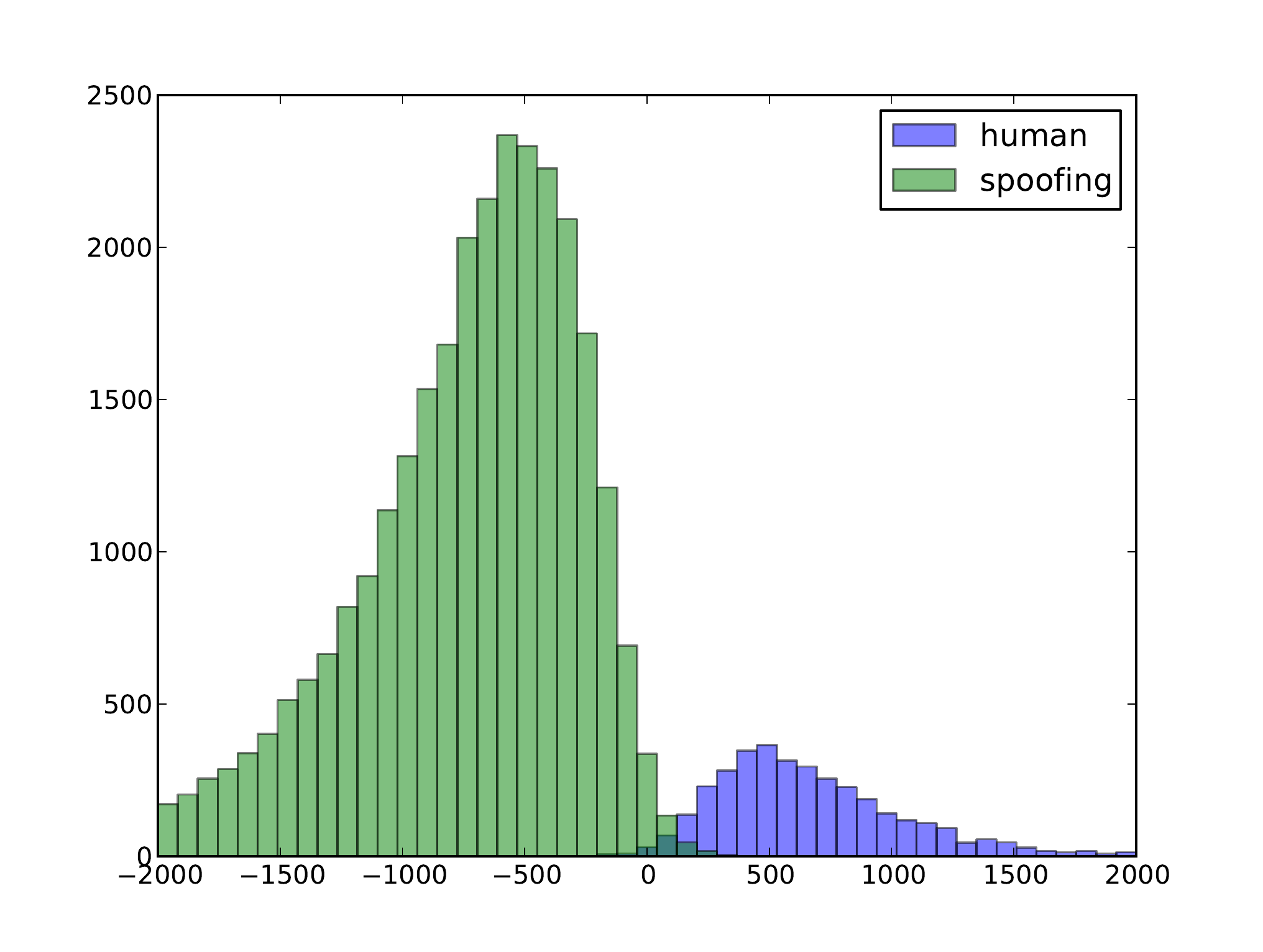}
    \caption{Scores distribution for spoofing (green) and authentic (blue) audios on the development set when using a GMM with 8 gaussians and bottleneck features}
    \label{fig:distribution}
  \end{figure}

    Figure~\ref{fig:distribution} shows the log-likelihood ratio (score) distribution obtained on the development set when a 8-mixture GMM was employed to classify the  bottleneck features. Score values vary in the interval $[-\infty,+\infty]$ and the higher the value, the higher the probability of the tested sample being authentic. The figure clearly shows this strategy provided a good separation over the develpment set. A similar behavior was verified for the SVM-RBF classifier.

    The third and last tested approach consisted of using the MLP trained for feature extraction directly as a classifier, without the removal of the output layer. In this case, the feature extraction was merged with the classification step.

    As the network last layer returns three values using the softmax function, according to presented in Figure~\ref{fig:mlp}, only $y_{0}$ is considered, since it represents the likelihood of being an authentic speech. Thus, values for this third approach vary in the interval $[0.0, 1.0]$. A score $(y_{0})$ was then calculated for each frame in the audio file, generating a score array for the entire audio. This array was used to compute a unique score for the audio sample. To do so, aiming at removing outliers within the audio file, the first 15\% lower array values are removed as well as the 25\% higher values. The remaining 60\% of the scores were then averaged, resulting in the final score.

    These three aproaches were, then, applied to the evaluation set, which contained samples comprising both known and unknown attacks. Results are presented in the next section.

  \section{Results and Analysis}
  \label{sec:results}

    Results obtained for the three tested systems are summarized in Table~\ref{tab:evalresults}. According to challenge rules, the adopted metric is the EER. For more details on what that means and how it is calculated, please refer to the contest evaluation plan \cite{wu2014asvspoof}.

\begin{table}[h]
\caption{\label{tab:mlp_results} {\it EER results (\%) obtained on development set and on evaluation set for known and unknown attacks.}}
\centerline{
\begin{tabular}{|c|c|c|c|c|}
\hline
\textbf{Classifier} & \textbf{Dev Set} & \textbf{Known} & \textbf{Unknown} & \textbf{All} \\ \hline
\textbf{SVM}        & 0.491            & 0.412          & 13.026           & 6.719        \\ \hline
\textbf{GMM}        & 0.658            & 0.443          & 12.796           & 6.620        \\ \hline
\textbf{MLP}        & 0.631            & 0.464          & 12.589           & 6.527        \\ \hline
\end{tabular}
}
\label{tab:evalresults}
\end{table}

    It can be seen that:

\begin{itemize}

\item the SVM-RBF classifier showed the best performance for known attacks, while the unknown attacks were better detected by the MLP classifier. However, EER values are very close, which means that the choice of the classifier is less determinant for the overall performance than the feature extraction mechanism itself.

\item all three systems performed very well for the known attacks, which shows that the network was successfull in capturing the pattern of attacks learned during training.

\item most of the unknown attacks were correctly detected; however a clear degradation of performance can be observed when error rates of known and unknown attacks are contrasted.

\item when considering only the five unknown attacks discriminated by method used (these results are not shown here due to space reasons), the proposed method obtained good results (EER near to 1\%) in three of them. Results for attacks S8 (a tensor-based voice conversion) and S10 (a speech synthesis algorithm implemented using the open source MaryTTS system), however, indicate a poor performance, with EERs of 26.8\% and 31.7\%, respectively.

\end{itemize}

    One hypothesis for the degradation observed in classifiers' performances for evaluation set is the occurence of overfitting to noise present in training samples. This situation can be verified by the existence of a significant difference in error rates even when training and testing samples are drawn from the same distribution. That is not what the results presented here show, since performance in the development set is close to the performance for known attacks in evaluation set.

    The second hypothesis is lack of generalization capacity, which means that some of the distinctive features learned by the network and the classifiers are not related to what distinguishes an authentic recording from spoofing attacks in general, but are rather due to patterns only observed in the known attack samples, i.e., specific characteristics of synthesis and conversion algorithms used during training step.

\begin{figure}[!htb]
    \centering
    \includegraphics[scale=0.75]{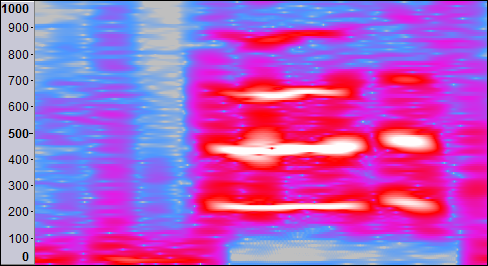}
    \caption{Low frequency noise discontinuity available on training and development set (0 to 1000 Hz in vertical axis)}
    \label{fig:lowfrequencyproblem}
  \end{figure}

    It was also verified after the submission that many spoofing audios available on the training and development sets present descontinuity in low frequency noise, mainly in the range 0 to 100 Hz. Figure~\ref{fig:lowfrequencyproblem} shows the problem. In this case, as 128 DCT  coefficients was used as DNN input, the first coefficients  will indicate this discontinuity and the network will learn this characteristic as a relevant feature to distinguish authentic from spoofing audios, degrading the network's generalization capacity when audios without this discontinuity are presented.

    Even though some degradation of performance is expected, the results obtained show that there is room for improvements, since the nature of unknown attacks is not inherently different from that of the known ones.

  \section{Conclusions}
  \label{sec:conclusions}

    The study presented here comprises the results obtained, along with the description of the systems implemented by CPqD for the Automatic Speaker Verification Spoofing and Countermeasures Challenge (ASVSpoof2015), held as a special session in INTERSPEECH 2015. The main goal of the challenge was the detection of spoofing attacks based on sinthesized and transformed speech.

    A speech feature extraction framework based on deep neural networks for spoofing detection is presented. The network can be used as a classifier itself or can be viewed as a bottleneck feature extractor feeding other classifiers. Two different classifiers were tested: a Gaussian Mixture Model and a Support Vector Machine with the radial basis function.

    The proposed systems were trained with the training set and tested on two different evaluation sets: one with attacks similar to those presented during training and another with unknown attacks, just as described in the evaluation plan.

    The use of a DNN as a feature extractor is of particular interest, as the generated features are fine-tuned to provide a good representation specifically for the problem to be solved, be it spoofing detection, speaker/speech recognition or other tasks. However, these features are highly dependent on the training samples and they can learn any bias present in this set. Thus the careful design of large and diverse datasets is even more relevant when using this kind of feature.

    Performance for the known attacks was satisfactory ($EER < 0.5\%$), indicating the adequacy of the proposed strategies. Results obtained for the unknown attacks were also promising. For some of the new attacks, however, the detection strategy had poor performance. This could be easily overcome with training data composed by samples generated by a more diverse attack techniques. In addition to an improved training set, the use of alternative forms of parametrization of the input audio in the neural network could be beneficial. Representations that make the speech phase spectrum more evident are specially interesting, as the use of such information proved to be highly successful in literature for spoofing detection \cite{wu2013synthetic}.

    Lastly, in future work, other network architectures, like Convolutional Neural Networks, should be tested in order to study which of them is able to provide better detection of unknown attacks, an ability extremely relevant in real-world applications, as rarely the techniques used by fraudsters for identity theft are known in advance.

  \section{Acknowledgements}

    We would like to gratefully thank MSc Christopher Dane Shulby for this paper review and helpful comments.

  \newpage
  \eightpt
  \bibliographystyle{IEEEtran}
  \bibliography{mybib}

\end{document}